\documentclass[preprint,12pt,showpacs,nofootinbib]{revtex4-1}

\pdfoutput=1

\usepackage{epsfig,color,graphicx}
\usepackage{amssymb,amsmath}
\usepackage{epstopdf}
\usepackage{url}

\usepackage{xspace}
\usepackage{epsfig}
\usepackage{amssymb,amsmath}

\def\be{\begin{equation}}
\def\ee{\end{equation}}
\def\bea{\begin{eqnarray}}
\def\eea{\end{eqnarray}}

\begin{document}

\begin{flushright}
ZU-TH-06/11 \\
\end{flushright}

\title{Collider Bounds on Lee-Wick Higgs Bosons}

\author{Ezequiel Alvarez}
\affiliation{CONICET, IFIBA, INFAP and Departamento de F\'isica, FCFMN, Universidad Nacional de San Luis, Av.~Ejercito de Los Andes 950, San Luis, Argentina}
\author{ Estefania Coluccio Leskow}
\affiliation{Departamento de F\'isica, FCEyN, Universidad de Buenos Aires, Ciudad Universitaria, Pab. 1, (1428) Buenos Aires, Argentina}
\author{ Jos\'{e} Zurita}
\affiliation{Institut f{\"u}r Theoretische Physik, Universit\"at Z\"urich, Winterthurerstrasse 190, CH-8057 Z\"urich, Switzerland. }

\begin{abstract}
We study the constraints on the Lee-Wick Higgs sector arising from direct collider searches. We work in an effective-field theory framework, where all of the Lee-Wick partners are integrated out, with the sole exception of the Lee-Wick Higgs bosons. The resulting theory is a two-Higgs doublet model where the second doublet has wrong-sign kinetic and mass terms. We include the bounds coming from direct Higgs searches at both LEP and Tevatron using the code {\tt {HiggsBounds}}, and show the currently excluded parameter space. We also analyze the prospects of LHC Run-I, finding that with a total integrated luminosity of $5~\rm{fb}^{-1}$ and a center-of-mass energy of 7 TeV, most of the parameter space for the SM-like CP-even Higgs will be probed.

\end{abstract}


\maketitle

\section{Introduction}

The Lee-Wick Standard Model (LWSM)~\cite{Grinstein:2007mp} is a recent proposal to solve the hierarchy problem.  The model is based on ideas of Lee and Wick~\cite{Lee:1969fy,Lee:1970iw} which were originally conceived for QED regularization in the late 60's.  In the most simple picture, the LWSM key feature consists in adding a Lee-Wick (LW) partner for each Standard Model (SM) field, but with larger mass and wrong sign kinetic term.  Henceforth, without modifying the low energy phenomenology, partial cancellations in loop corrections eliminate quadratic divergences in the Higgs mass.

One of the main controversies in the Lee-Wick ideas is, at the same time, one of the most challenging and attractive points of the theory: the LW-partners fields seem to violate microscopic causality.  Despite the alarm this statement may provoke at a first glance, the theory has remained robust during these last years.  As a matter of fact, the theory respects macroscopic causality, since LW resonances are only intermediate unstable states and therefore the resulting theory is unitary at least at tree level~\cite{Grinstein:2007mp}.  Moreover, it has been proved that a ${\cal O}(N)$ Lee-Wick theory is unitary at all orders~\cite{Grinstein:2008bg}, and it has also been shown that violation of microscopic causality does not imply any paradox as far as this violation is microscopic enough~\cite{Coleman:1969xz}.  It is fair also to recall at this point that there has never been found any inconsistency in the original Lee-Wick ideas, but, instead, they lost the due attention because of the seminal paper of 't Hooft and Veltman which finally solved the regularization and renormalization of gauge theories~\cite{'tHooft:1972fi}.

Several phenomenological and theoretical aspects of the LWSM have been studied in the last few years~\cite{Grinstein:2007iz,Fodor:2007fn,Knechtli:2007ea,Grinstein:2008qq,FCNC,Wu:2007yd,Espinosa:2007ny,Rizzo:2007ae,Krauss:2007bz,Carone:2008bs,Rizzo:2007nf,Carone:2008iw,Carone:2009it,Alvarez:2008za,Alvarez:2008ks,Underwood:2008cr,Grinstein:2008bg,Fornal:2009xc,Carone:2009nu,Chivukula:2010nw,Chivukula:2010kx,Espinosa:2011js}. In particular, electroweak precision measurements \cite{Carone:2008bs,Alvarez:2008za,Alvarez:2008ks,Underwood:2008cr,Chivukula:2010nw} force the LW gauge bosons and quark masses (assuming Minimal Flavor Violation) to be around a few TeV. The LW leptons are only constrained by direct detection through pair production, and thus the bound is about $100~\rm{GeV}$ \cite{PDG}.  

On the other hand, the scalar sector of the theory has been indirectly constrained in a practical and efficient article by Carone and Primulando \cite{Carone:2009nu}.  In this paper the authors obtain bounds on the LW Higgs sector through the one-loop contributions of Higgs bosons to processes such as $B$-meson mixing, $b\to X_s \gamma$ and $Z\to b \bar b$. Despite being formally one-loop suppressed, these kind of observables are enhanced by large Yukawa couplings and are very well measured.  Their combined analysis yields a bound to the non SM-like Lee-Wick Higgs scale of around $400$ GeV.

In this article, as a complement of the indirect analysis performed in Ref.~\cite{Carone:2009nu}, we study the constraints to the LW Higgs sector arising from direct searches at LEP, Tevatron and LHC.  Although this work achieves better present bounds than the indirect study only in a minor part of the phase space, it settles the perspectives for direct detection of the LW Higgs sector.  Moreover, we show that after LHC run I, the non SM-like Higgs parameter space should be mostly probed up to the TeV scale, which will improve considerably the present indirect limits.

On a first step, we implement the present collider direct constraints from LEP and Tevatron with the help of the code {\tt {HiggsBounds}} \cite{Bechtle:2008jh,Bechtle:2009ic,Bechtle:2009ni,Bechtle:2010jt,Bechtle:2011sb}, and study the exclusion of the parameter space.  This computer code allows to analyze simultaneously several LEP and Tevatron direct search signals and choose the best one to constrain a given point in parameter space at a $95\ \%$ C.L. .  In a second step,  we include the exclusion projections from the ATLAS collaboration, finding that the LHC with a center-of-mass energy of 7 TeV and an integrated luminosity of 5 $\rm{fb}^{-1}$ (expected to be achieved by the end of 2012) would be able to probe most of the parameter space for the lightest Higgs mass. We also focus on the heavy Higgs bosons, showing the prospect for their detectability at the LHC Run I.

This paper is organized as follows: In Section~\ref{sec:intro} we review the LWSM Higgs sector, focusing on the couplings and relevant branching fractions of the Higgs bosons. We summarize and analyze the input needed to run {\tt {HiggsBounds}}.  In Section~\ref{sec:num} we obtain the main direct bounds to the Higgs sector coming from LEP and Tevatron signal analysis.  Afterwards, we study the LHC perspectives for direct search in the $WW$ and $\gamma \gamma$ channels.  Section~\ref{sec:conclu} contains our conclusions and final remarks.

\section{The Lee-Wick Standard Model Higgs sector}
\label{sec:intro}
In this section we write down and discuss the LWSM Higgs sector in the physical basis and compute all the collider-relevant effective couplings which involve scalars.  These effective couplings are used in next section to constrain the parameter space of the model using all the suitable results from LEP and Tevatron with the aid of {\tt {HiggsBounds}}. 

\subsection{Theoretical setup}
\label{sec:setup}
As mentioned in the Introduction, the LW Lagrangian can be written in two equivalent
 ways. One of them introduces none additional fields, but involves higher derivatives (HD) which give additional degrees of freedom to the theory. The other one introduces the new degrees of freedom explicitly as new partner fields for each SM field. This second formulation --the LW picture-- is more clear in dealing with collider experiments, since the contribution of these new fields to the processes can be understood in terms of the usual Feynman diagrams.  Along this work the LW formulation will be employed.

We now focus on the LWSM Higgs sector Lagrangian.  (The Lagrangian and details of the full LWSM can be found elsewhere \cite{Grinstein:2007mp,Alvarez:2008ks,Underwood:2008cr}.)  Given the stringent bounds on the LW gauge bosons, $M_1,M_2 \gtrsim 2.4$ TeV, and the quark masses $\gtrsim 4$ TeV found in previous electroweak precision tests studies \cite{Underwood:2008cr,Alvarez:2008ks,Chivukula:2010nw}, and given the relatively loose bounds on the charged Higgs, $m_{\tilde h^\pm} \gtrsim 450$ GeV, coming from $b \to s \gamma$ bounds \cite{Carone:2009nu}, it is suitable to decouple from the theory the LW gauge bosons and quarks and to keep only the LW scalars.  Notice that although the LW leptons cannot be constrained from EWPT and their only mass limit comes from direct search, $m_{leptons} > 100$ GeV \cite{PDG},  they can be safely decoupled due to their tiny Yukawa couplings to the Higgs sector.  We may also neglect mixing between SM and LW leptons.   Within these assumptions the LWSM Higgs sector reads before symmetry breaking and in the flavor basis,
\bea\label{eq:Lag}
{\cal L}_{Higgs} &=& (D_{\mu} H)^\dagger D^{\mu} H - (D_{\mu} \tilde{H})^\dagger D^{\mu} \tilde{H} + M^2 \tilde{H}^\dagger \tilde{H} - V (H- \tilde{H}) + \nonumber \\ 
&&\left( g_u^{ij } \bar{u}^i_R (H-\tilde H) \epsilon Q^j_L - g_d^{ij } \bar{d}^i_R (H^{\dagger}-\tilde H^\dagger) Q^j_L - g_e^{ij} \bar{e}^i_R (H^{\dagger}-\tilde H^\dagger ) L^j_L  + \rm{h. c.} \right) , 
\eea
where 
\bea
V(X)= - \frac{m^2}{2} X^\dagger X + \frac{\lambda}{4} (X^{\dagger} X)^2 \, ,
\eea
the $Q^j_L$ and $E^j_L$ are the $SU(2)_L$ doublets for the $j^{th}$ family of quarks and leptons, respectively, and $g_X^{ij}$ the corresponding Yukawa coupling before diagonalization.  Since we have decoupled the LW gauge bosons from the theory, $D_{\mu}$ is the usual SM $SU(2) \times U(1)$ covariant derivative.

The Lagrangian ${\cal L}_{Higgs}$ corresponds to a two-Higgs doublet model (2HDM) which is different from the usually found in the literature, mainly due to the opposite sign in the kinetic terms of the second Higgs.  Notice also that contrary to the usual type I and type II models, here both Higgs bosons couple to all fermions.  However, as first observed in Ref.~\cite{Carone:2009nu}, after electroweak symmetry breaking it is possible --and useful-- to identify some pieces of this 2HDM Lagrangian with the MSSM model for different values of $\tan \beta$.  

The diagonalization to the physical basis is driven in two steps after electroweak symmetry breaking. First, one diagonalizes the scalar fields in the quadratic part of the kinetic plus potential Lagrangian (first line in Eq.~(\ref{eq:Lag})), and second, the fermion basis is diagonalized in the Yukawa interactions (second line in Eq.~(\ref{eq:Lag})).

Spontaneous symmetry breaking is understood as usual through the Higgs potential (last two terms in the first line of Eq.~(\ref{eq:Lag})).  In this case, the minimization conditions forbid $\tilde H$ to acquire a vacuum expectation value, leaving the usual SM relation $m^2=\lambda v^2 / 2$ , where $v \approx 246~{\rm GeV}$ is the vacuum expectation value of $H$. The expressions for the electroweak gauge boson masses are exactly the same as in the SM.  In the unitary gauge one can write
\be
 H=\frac{1}{\sqrt{2}}
\left( \begin{array}{c}   0   \\   v+h   \end{array}  \right) 
\, , \qquad 
\tilde{H}=\frac{1}{\sqrt{2}}
\left( \begin{array}{c}  \sqrt{2} \,  \tilde{h}^{+}   \\
   \tilde{h}+ i \tilde{P}      
\end{array}
\right)  \, .
\ee
The spectrum consists of five Higgs fields: two neutral CP-even Higgs, ($h$ and $\tilde{h}$), one neutral CP-odd Higgs $\tilde{P}$ and two charged bosons, $\tilde{h}^{\pm}$.   

After electroweak symmetry breaking one can diagonalize the scalar sector.  The neutral CP-even bosons mix with each other via a symplectic transformation
\be\label{eq:eigenstates}
\left(
\begin{array}{c}
  h      \\
  \tilde{h}       
\end{array}
\right)
=
\left(
\begin{array}{cc}
  \cosh \theta&   \sinh \theta   \\
 \sinh \theta & \cosh \theta       
\end{array}
\right)
\left(
\begin{array}{c}
  h_0      \\
  \tilde{h}_0      
\end{array}
\right) \, ,
\ee
where the subscript $0$ is reserved for the mass eigenstates. The mass eigenvalues are given by
\be
m^2_{h_0,\tilde{h}_0}=\frac{M^2}{2} \bigl( 1 \mp \sqrt{1-\frac{4m^2}{M^2}} \bigr) \, .
\ee
and the mixing angle factors by
\be
\cosh \theta = \frac{1}{(1-r^4)^{1/2}} \, , \qquad \sinh \theta = \frac{ - r^2}{(1-r^4)^{1/2}} \, , \qquad r \equiv \frac{m_{h_0}}{m_{\tilde{h}_0}} \, .
\ee
At tree level, both $\tilde{P}$ and $\tilde{h}^{\pm}$ are degenerate in mass, and the following sum rule holds
\be\label{eq:mhsumrule}
m^2_{h_0} + m^2_{\tilde{h}_0}=m^2_{\tilde{P}} = m^2_{\tilde{h}^{\pm}}=M^2 \, .
\ee
Note that the model implies the following hierarchy in the spectrum: $m_{h_0} \le m_{\tilde{h}_0} < m_{\tilde{P}}$.

In order to diagonalize the Yukawa interactions in Eq.~(\ref{eq:Lag}), since we have decoupled LW fermions, the usual SM rotation in the quark fields shall be performed.  Henceforth, the resulting Yukawa fermion-scalars interacting Lagrangian may be divided in a neutral scalar and charged scalar Lagrangian.  The neutral scalar Lagrangian is as the usual SM Lagrangian replacing 
\be
\label{eq:replace}
h\to h-(\tilde h + i \tilde P).
\ee
The charged scalar Lagrangian for the quark fields is
\be\label{eq:Lhpmff}
{\cal L}_{\tilde{h}^\pm f \bar{f}} = \frac{\sqrt{2}}{v} 
\left[ \tilde h^+ \left( \bar{u}_R M_u V d_L - \bar{u}_L M_d V d_R \right) 
+ \tilde h^- \left( - \bar{d}_R V^{\dagger} M_d u_L + \bar{d}_L V^{\dagger} M_u u_R \right),
\right]  \, ,
\ee
where $V$ is the SM CKM matrix and $M_{u,d}$ are the diagonal mass matrices.  Notice that Eq.~(\ref{eq:Lhpmff}) is up to an overall minus sign, 
 since $\tilde{h}^{\pm}$ comes from $\tilde{H}$ doublet, the same result that one obtains in a general type II 2HDM for $\tan \beta=1$.  To obtain the physical basis interaction $h$ and $\tilde h$ shall be rotated to their physical counterparts $h_0$ and $\tilde h_0$ through Eq.~(\ref{eq:eigenstates}).

Having diagonalized the Higgs sector we now write down the relevant effective couplings of these scalars to fermions, to gauge bosons, and between them.  The customary notation is such that $g_{\phi X Y}$ denotes the effective coupling of any Higgs boson $\phi$ to the fields specified in the string $XY$, normalized to the SM case for neutral Higgs bosons, and to charged would-be Goldstone bosons for the charged Higgs.  If the process does not have an analog in the SM, then the normalization will be defined in each case.

\subsubsection{The Kinetic and potential couplings}

When the kinetic and potential terms (first line in Eq.~(\ref{eq:Lag})) are rotated to the physical basis, gauge bosons to scalar and trilinear scalar couplings, among others, show up.

We first turn our attention to the interactions between scalars and gauge bosons. Since quartic interactions are not relevant for the LHC collider phenomenology, we are left to consider trilinear couplings.
One has to keep in mind that in the interaction basis only $H$ is able to couple to a vector boson pair since $v \ne 0$
. Using $h=\cosh \theta h_0 + \sinh \theta \tilde{h}_0$, one gets that the ratio of the LWSM coupling to the usual SM coupling is
\begin{equation}\label{eq:hvv}
g_{h_{0} V V} = \cosh \theta \, , \qquad g_{\tilde{h}_0 V V} = \sinh \theta \, .
\end{equation}
The two other relevant interaction are $x \tilde{P} Z$ and $x \tilde{h}^{\pm} W^{\mp}$, where $x$ is a CP-even Higgs boson. They come entirely from $\tilde{H}$, and therefore one gets
\be\label{eq:hhv}
g_{h_0 \tilde{P} Z} = g_{h_0 \tilde{h}^{\pm} W^{\mp}}  = - \sinh \theta \, , \qquad  g_{\tilde{h}_0 \tilde{P} Z}= g_{\tilde{h}_0 \tilde{h}^{\pm} W^{\mp}}= - \cosh \theta \, ,
\ee
where, in the absence of a SM reference process, we have normalized to $g_2 / c_W = e / c_W s_W$. Given the hierarchy of the mass spectrum, the only kinematically allowed decay modes involving these couplings are $\tilde{P} \to h_0 / \tilde{h_0} Z$ and $\tilde{h}^\pm \to h_0 / \tilde{h_0} W^\pm$. The corresponding decay widths read
\be
 \Gamma (\tilde{A} \to V x)  = \frac{G_F}{8 \sqrt{2} \pi m^3_{\tilde{A}} }  \, g^2_{x \tilde{A} V} \, \lambda^{3/2} (m_x,m_V,m_{\tilde{A}}) \, ,
\ee
where the kinematic factor $\lambda$ is given by
\be\label{eq:lamJ}
\lambda (m_1,m_2,m_3) = (m_1^2+m_2^2-m_3^2)^2 - 4 m_1^2 m_2^2 \, .
\ee
and $\tilde{A}=\tilde{P},\tilde{h^\pm}$, $V=Z,W^{\pm}$ and $x=h_0, \tilde{h_0}$.

For the trilinear Higgs interactions, all the couplings depend only on the single factor
\begin{equation}\label{eq:Ctilde}
%\tilde{C} = 
\frac{\lambda v}{2}  = \frac{1}{(1+r^2)} \, \frac{m^2_{h_0}}{v} = \frac{r^2}{(1+r^2)} \, \frac{m^2_{\tilde{h}_0}}{v} \, .
\end{equation}
The final expression for the absolute trilinear coupling of three generic Higgs bosons $xyz$ is given by
\be\label{eq:hhh}
g_{xyz} = - \frac{\lambda v}{4}  c\,  g_{x f \bar{f}} g_{y f \bar{f}} g_{z f \bar{f}} \, ,
\ee
where $c=2$ if there are charged Higgs bosons, $c=3$ for $h_0 h_0 \tilde h_0$ and $h_0 \tilde h_0 \tilde h_0$, $c=1$ for all other cases, and $g_{xf\bar f}$ are given below. The CKM matrix shall be set to unity in this formula.  For instance, $g_{h_0 \tilde h^+ \tilde h^-}=\frac{\lambda v}{2} (\sinh \theta -\cosh \theta)$. The only possible decay involving three Higgs bosons, due to the hierarchy in the spectrum and the CP quantum numbers, is\footnote{However, the $h_0 \tilde{h}^+ \tilde{h}^-$  and $\tilde{h}_0 \tilde{h}^+ \tilde{h}^-$couplings are a necessary ingredient for the decay of neutral Higgs bosons into $\gamma \gamma$ and $\gamma Z$, as will be explained in Subsection \ref{subsec:loopdec}.  } $\tilde{h}_0 \to h_0 h_0$. The decay width in this channel reads 
\be
 \Gamma (\tilde{h}_0 \to h_0 h_0)  = \frac{1}{8 \pi m_{\tilde{h}_0}} \sqrt{1-4 r^2} \, \, \, g_{\tilde{h}_0  h_0  h_0}^2 \, , \qquad \text{with}  \quad g_{\tilde{h}_0  h_0  h_0} =  3 \frac{\lambda \,v}{4} \, g_{h_0 f \bar{f}}^3 \, ,
 \ee 
where the corresponding symmetry factors due to the indistinguishability of the $h_0$ bosons have been added.

Quartic Higgs interactions are not relevant for collider phenomenology.

\subsubsection{The Yukawa couplings}

The Yukawa couplings for the neutral Higgs are easily derived by performing the replacement of Eq.~(\ref{eq:replace}) in the usual SM Lagrangian and then rotating the fields through Eq.~(\ref{eq:eigenstates}). One gets
\begin{equation}
\label{eq:hff}
g_{h_{0} f \bar{f}} = - g_{\tilde{h}_0 f \bar{f}} = \cosh \theta - \sinh \theta = \frac{1+r^2}{\sqrt{1-r^4}}  \, , \qquad g_{\tilde{P} f \bar{f}}  =  -1 \, .
\end{equation}

In the charged Yukawa interactions the most relevant coupling occurs in the third generation where the new channel $t\to\tilde h^+ b$ opens for the top quark decay in case of a light Higgs.  Neglecting the bottom mass, the width for this decay is
\be
\Gamma (t \to \tilde{h}^{+} b) =\frac{G_F}{8 \sqrt{2} \pi} m_t^3 (1-\frac{m_{\tilde{h}^\pm}^2}{m_t^2})^2\, .
\ee

\subsubsection{The loop-mediated effective couplings}
\label{subsec:loopdec}

The aim of this subsection is to compute the relevant information concerning the loop-mediated effective couplings of Lee-Wick Higgs bosons to two gluons, two photons, and a photon and a $Z$. 
Effective couplings of $h_0$ to gluons and photons were already presented in \cite{Krauss:2007bz}. 

\begin{itemize}
\item {\it Effective coupling of Higgs and two gluons}

The gluonic case is straightforward, because it involves a single diagram: the triangle fermion loop. Since we are decoupling the LW top quark and the other fermions may be ignored due to their suppressed Yukawa coupling, the only fermion we keep in the loop is the top. Therefore, the only modification with respect to the SM case is simply the change in the top Yukawa coupling, which can be read from Eq.~(\ref{eq:hff}). Hence, $g_{h_0 gg}=g_{h_0 f\bar{f}} = - g_{\tilde{h}_0 gg} = - g_{\tilde{h}_0 f \bar{f}}$.

For $\tilde{P}$ the loop function for fermions is different with respect to the one for CP-even Higgs bosons (see Appendix \ref{app:loopfunc}) due to the CP-odd nature of $\tilde{P}$. Therefore one has
\be\label{eq:Pgg}
g_{\tilde{P}gg}^2=\frac{\sigma (gg \to \tilde{P}) }{\sigma^{SM} (gg \to H) } = \lvert \frac{g_{\tilde P t \bar t}\, F_{1/2}^{\tilde P}(\beta^t_{\tilde P})}{F_{1/2}(\beta^t_{\tilde P})}  \rvert^2 \, ,
\ee
where $\beta^t_{\tilde{P}}=(2 m_t / m_{\tilde{P}})^2$, the SM cross-section is evaluated at $m_H=m_{\tilde{P}}$, and the loop functions are defined in Appendix \ref{app:loopfunc}.  

\item {\it Effective coupling of Higgs and two photons}

In the photon case, in addition to the top quark loop discussed for the gluon case, one has to consider also $W^{\pm}$ and charged Higgs in the loops, including the proper relative scaling factors for each amplitude. 

For the top amplitude, one has to multiply the loop function $F_{1/2}$ by $N_c  Q_{top}^2$ and by the relative coupling $g_{xt\bar{t}}$, where $x=h_0,\tilde{h_0},\tilde{P}$. The $W$ loop amplitude has to be simply multiplied by $g_{xVV}$. Finally, the charged Higgs loop contribution will include the $F_0$ function, an extra factor of $m_W^2/m_{\tilde{h}^\pm}^2$
, and  the $g_{x \tilde{h}^+ \tilde{h}^-}$ coupling.  This yields
\begin{equation}
\label{eq:hgaga}
g_{x \gamma \gamma} = \frac{g_{x t \bar{t}} N_c Q_t^2 F^x_{1/2} (\beta^t_x) +g_{xVV} F_1(\beta^W_x) 
+ \frac{g_{x \tilde h^+ \tilde h^-}}{2 m_W^2/v} \, \frac{m_W^2}{m_{\tilde{h}^{\pm}}^2} F_{0} (\beta^{h^{\pm}}_x)  }
{N_c Q_t^2 F_{1/2} (\beta^t_x)+F_1(\beta^W_x)} \, ,
\end{equation}
where $N_c=3$, $Q_t=2/3$ and $F^{h_0,\tilde h_0}_{1/2}=F_{1/2}$.   Notice that since $g_{\tilde P VV}=g_{\tilde P \tilde h^+ \tilde h^-}=0$ then only the top loop contributes to the numerator of the $g_{\tilde P \gamma \gamma}$ coupling.  Our expression for $g_{h_0 \gamma \gamma}$ agrees with the ones in Refs.~\cite{Krauss:2007bz} and \cite{Cacciapaglia:2009ky}\footnote{These two groups do not agree between them at the NLO level, however, since we are decoupling the LW partners and working at LO, we do agree with both of them at this level.}.

\item {\it Effective coupling of Higgs to Z and photon}

The last of the loop-mediated interactions is the decay of a neutral Higgs boson into a photon and a $Z$ boson. The corresponding MSSM formulae can be found in Appendix C. of Ref. \cite{Gunion:1989we}. Since this process can be obtained simply by replacing a photon by a $Z$ boson with respect to the $\gamma \gamma$ channel, from our previous discussion one already knows all the required relative factors. However, it is worth noting that the presence of the $Z$ yields rather cumbersome expressions which read
\be
g_{x Z \gamma}  =
\frac{g_{x t \bar{t}} A^x_t ( \beta_x^t,\beta_Z^t)+g_{xVV} A_W (\beta_x^W, \beta^W_Z) +
\frac{g_{x \tilde h^+ \tilde h^-}}{2 m_W^2/v} A_{\tilde h^{\pm}} (\beta_x^{h^{\pm}},\beta_Z^{h^{\pm}})  }
{ A_t^{h_0} (\beta^t_{x},\beta^t_Z) + A_W  (\beta_x^W, \beta_Z^W)} \,. 
\ee 
Again, notice that for $\tilde P$ only the top loop contributes in the numerator.

\end{itemize}

\subsection{Phenomenological qualitative analysis}
\label{subsec:phenoanal}

In this section we analyze some qualitative features of the LWSM Higgs sector above presented in order to understand how the different Higgs bosons will face the direct searches performed by LEP and Tevatron.  This analytic discussion is complemented by the quantitative numeric analysis performed in next sections by the implementation of {\tt {HiggsBounds}} 2.1.0 .

\begin{figure}[!htbp]
\begin{center}
\begin{minipage}[b]{0.45\linewidth}
\begin{center}
\includegraphics[width=1\textwidth]{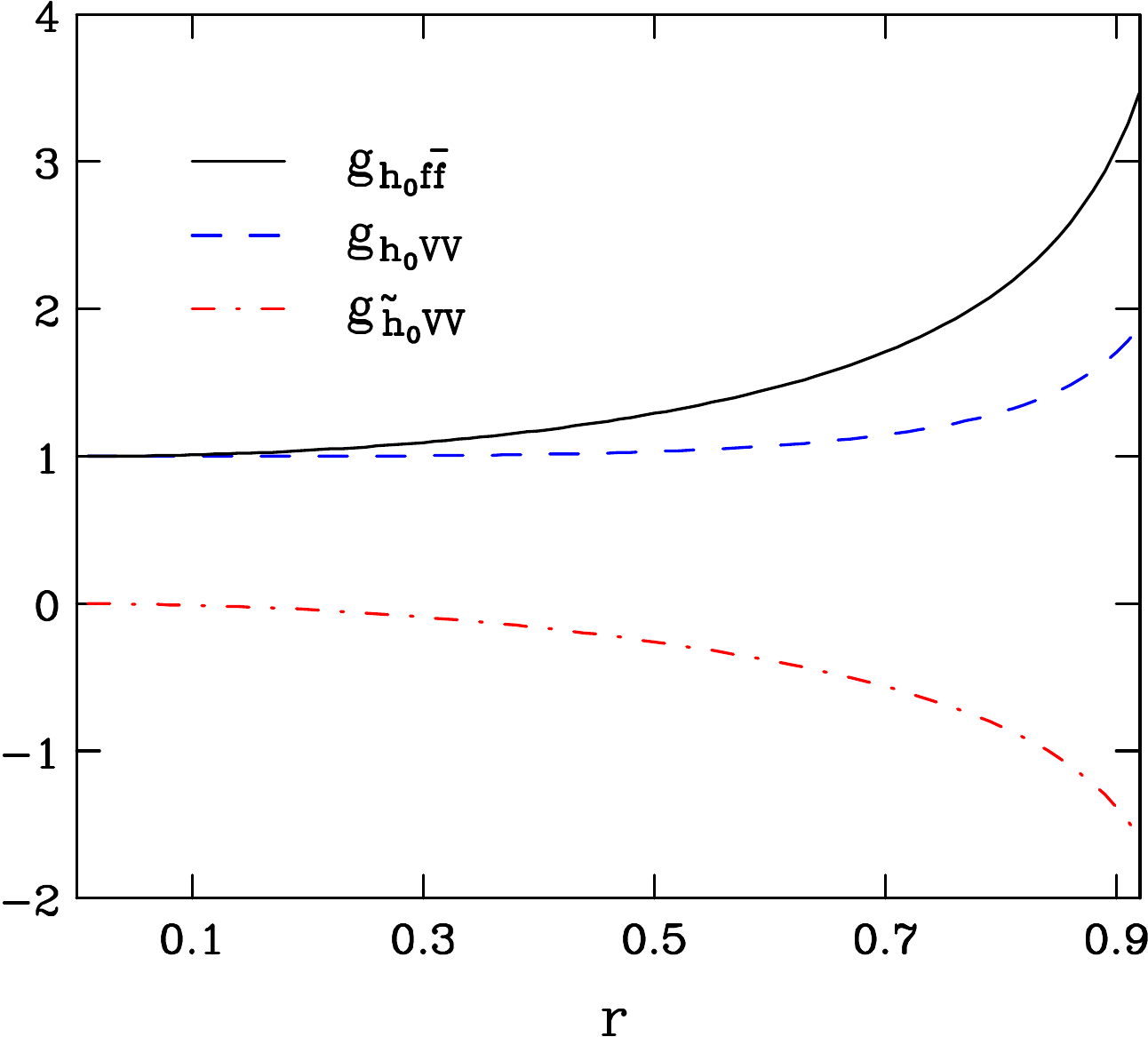}
\newline
(a)
\end{center}
\end{minipage}
\hspace{0.2cm}
\begin{minipage}[b]{0.45\linewidth}
\begin{center}
\includegraphics[width=1.08 \textwidth]{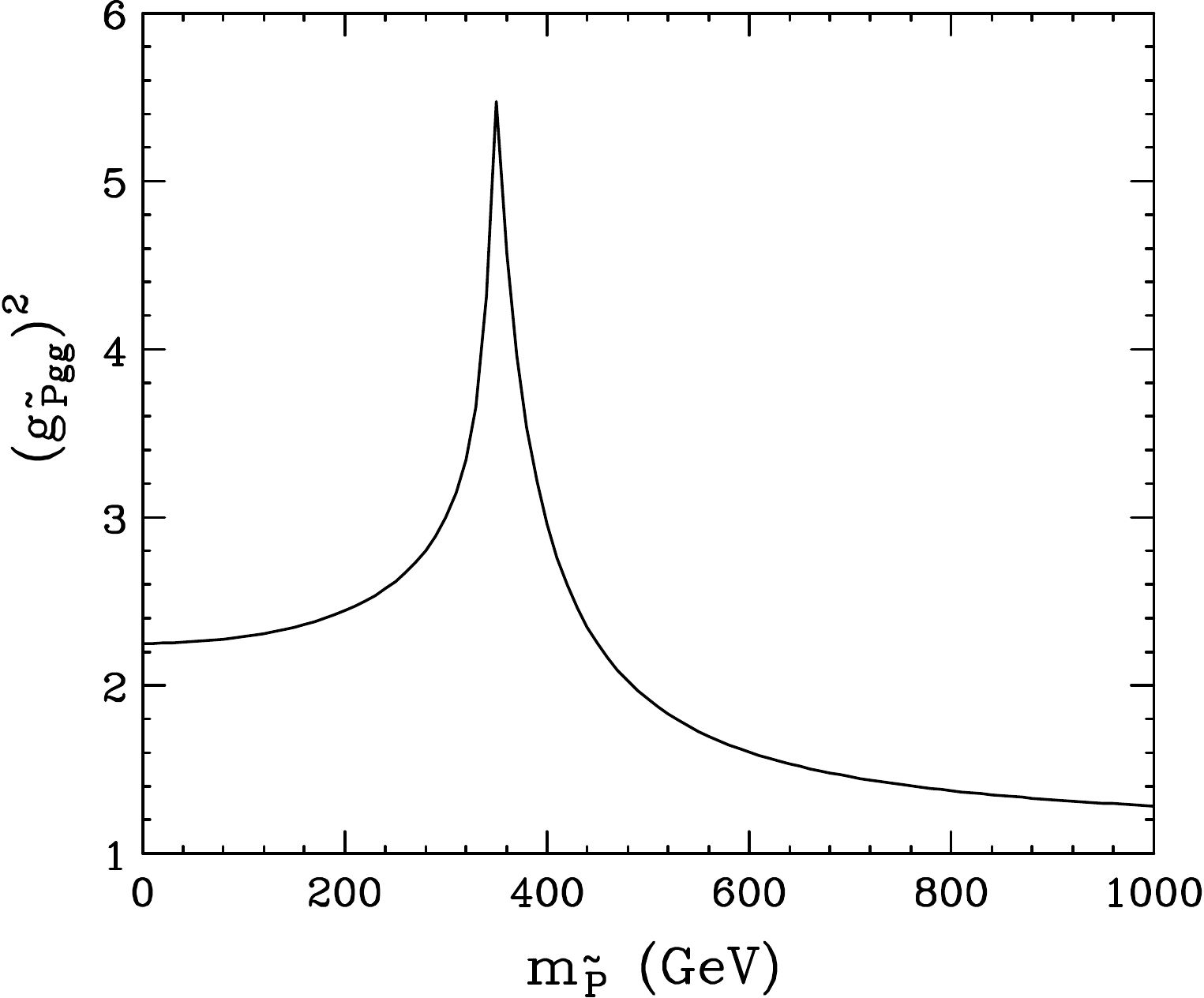}
\newline
(b)
\end{center}
\end{minipage}
\caption{ (a) From top to bottom, relative couplings of the neutral SM-like Higgs to fermions $g_{h_0 f \bar f}$ (which is  equal to $g_{h_0 gg}=-g_{\tilde h_0 gg}=-g_{\tilde h_0 f \bar f}$), to gauge bosons $g_{h_0 VV}$, and of the LW CP-even Higgs to gauge bosons $g_{\tilde h_0 VV}$, as a function of the ratio of the physical masses, $r$.  Notice the increase in the couplings for $r \to 1$, while for $r \to 0$, $h_0$ behaves as the SM Higgs.
We also note that $g_{h_0 f\bar f} > g_{h_0VV}$ and $g_{\tilde{h}_0VV} ^2 < g_{h_0VV}^2$ hold. (b)  $g_{gg \tilde{P}}^2$ as a function of $m_{\tilde{P}}$ in GeV. Notice that the $\sigma(gg\to \tilde P)$ cross section is always greater than the corresponding value for a SM (CP-even) Higgs of the same mass, and that this relative coupling peaks at the value $2 m_t$.}
\label{fig:1}
\end{center}
\end{figure}

As a first step it is interesting to plot  (Fig.~\ref{fig:1}a) the relative couplings $g_{h_0 VV},\ g_{\tilde h_0 VV}$ and $g_{h_0 f\bar f}$ (which is equal to $g_{h_0 gg}=-g_{\tilde h_0 gg}=-g_{\tilde h_0 f \bar f}$) as a function of their only variable, $r$.  Both $g_{h_0 gg}^2$ and $g_{h_0 VV}^2$ are greater than one, and this has direct implications in collider physics bounds: at LO, the SM-like Higgs production cross sections in all relevant channels at LEP, Tevatron and LHC (gluon fusion, vector boson fusion, Higgs-strahlung, associated production with gauge bosons and/or heavy quarks, bottom fusion) are always larger that the SM ones. In particular, this implies that the LEP $114.4~\rm{GeV}$ bound on the SM Higgs would also apply to $h_0$, as argued in Ref.~\cite{Carone:2009nu}.

The effective couplings in Fig.~\ref{fig:1}a are monotonic functions of the variable $r$. Moreover, their values blow in the limit $r \to 1$. However, in this limit one enters in the non perturbative regime: the strongest constrain comes from requiring that the top Yukawa coupling squared is smaller than $4 \pi$, which yields $r < 0.923$. For this extreme value of $r$, the effective couplings are $g_{h_0 f \bar{f}} = 3.54$,  $g_{h_0 VV}=1.91$ and $g_{\tilde{h}_0 VV}=-1.63$. 

At this point, it is worth noticing an important peculiarity of the LW Higgs sector: the sum rule for the gauge boson couplings to CP-even Higgs is
\be
g_{h_0 VV}^2 - g_{\tilde{h}_0 VV}^2 =1 \, ,
\ee
while the general 2HDM result involves a $+$ sign. This means that in the LW model \emph{both} CP-even Higgs bosons can have sizable couplings to the gauge bosons (and even be larger than one), in contrast with the most general 2HDM result. In the limit $r \to 0$, the LW Higgs doublet $\tilde{H}$ decouples. Hence, it is not a surprise to see that, in this region, the $h_0$ couplings tend to the SM value, while $\tilde{h}_0$ is gaugephobic.

Another interesting point to note is that the following inequality holds
\be\label{eq:effcouphierarchy}
g_{h_0gg}^2 > g^2_{h_0 VV} > g^2_{\tilde{h_0} VV}   \, .
\ee
This result has a strong implication for hadron-collider phenomenology. At the Tevatron and the LHC, the different production modes will scale with either the fermion/gluon (gluon fusion, associated top production, bottom fusion) or the vector boson coupling (VBF, Higgstrahlung, etc).  Given that the gluon fusion mode has the largest cross section, this implies that the effective-gluon coupling squared is actually an upper bound for the enhancement factor of the \emph{total} production cross section.

In Fig.~\ref{fig:1}b we show the relative effective coupling of the CP-odd LW Higgs to gluons as a function of $m_{\tilde{P}}$. We see that, as expected, the relative cross section peaks at $m_{\tilde{P}}=2 m_t$.   Although we do scan this region in parameter space in next section, this peak is already outside of the allowed region in Ref.~\cite{Carone:2009nu}.  We find that values for $g_{\tilde P gg}^2$ are $4.6 \, (1.6)$ for $m_{\tilde{P}}=463 \, (1000)~\rm{GeV}$. 

\begin{figure}[!htbp]
\begin{center}
\begin{minipage}[b]{0.45\linewidth}
\begin{center}
\includegraphics[width=1\textwidth]{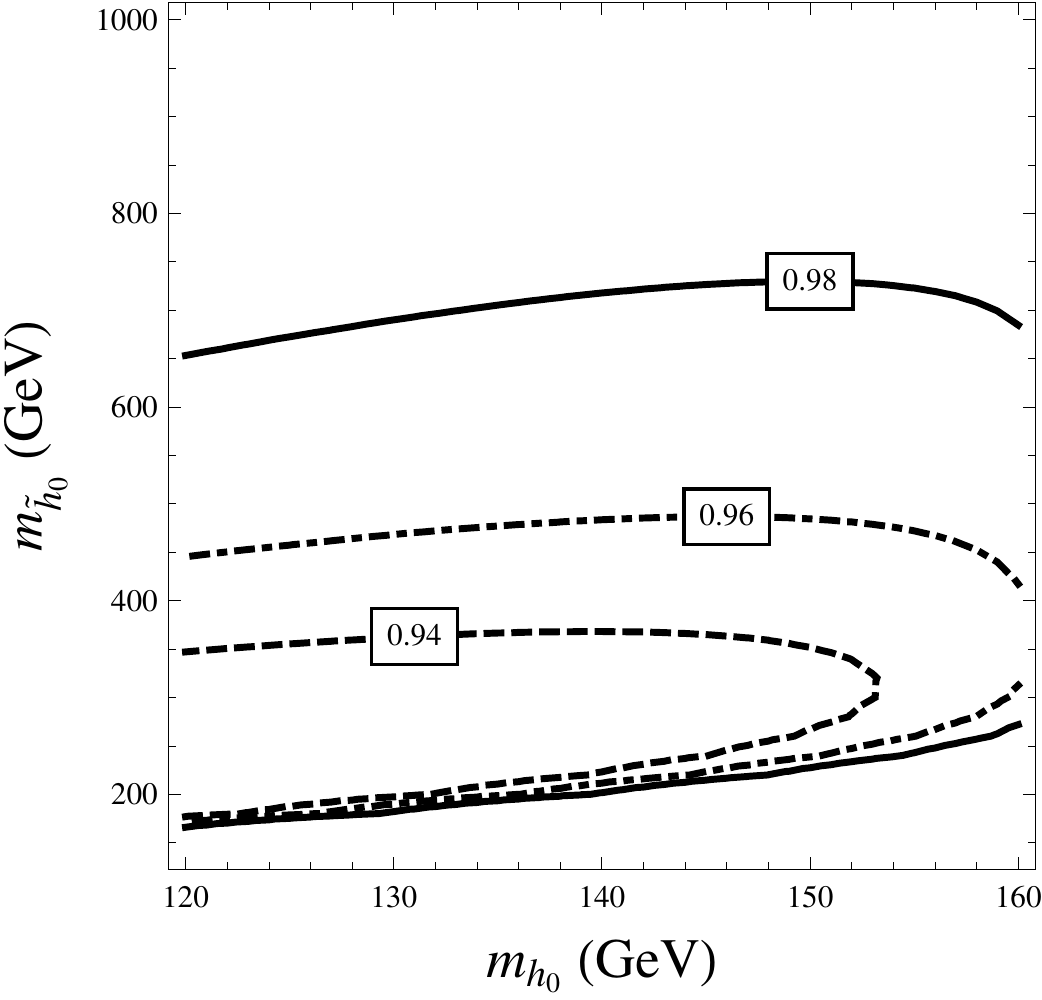}
\newline
(a)
\end{center}
\end{minipage}
\hspace{0.5cm}
\begin{minipage}[b]{0.45\linewidth}
\begin{center}
\includegraphics[width=1\textwidth]{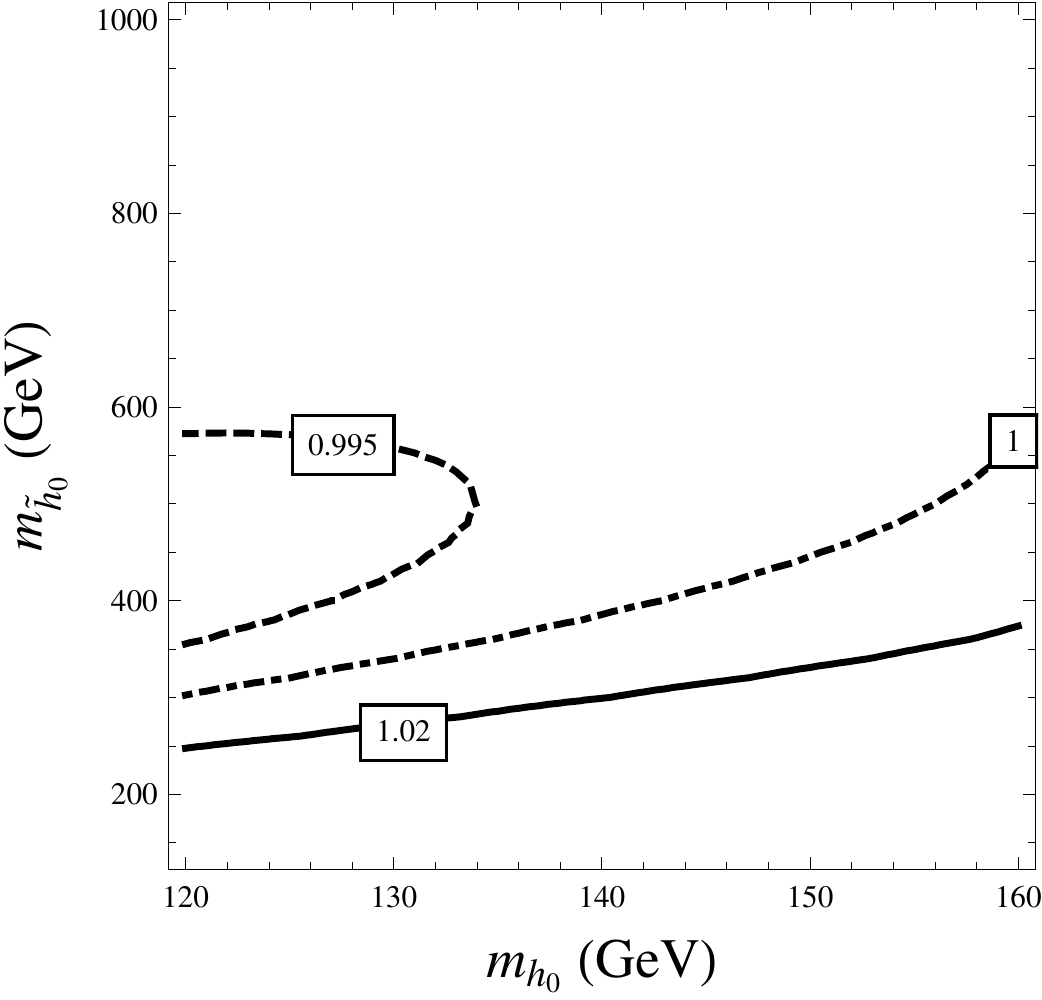}
\newline
(b)
\end{center}
\end{minipage}
\caption{Contour plots of  $\lvert g_{h_0 \gamma \gamma} \rvert^2$ (a) and $\lvert g_{h_0 Z \gamma} \rvert^2$ (b) in the $m_{h_0},m_{\tilde{h}_0}$ plane. Notice that both quantities are generally below $1$. For the $\gamma \gamma$ case, one finds that in the region of parameter space not excluded by the analysis of Ref.~\cite{Carone:2009nu}, the relative coupling squared is strictly below one, while for the $Z \gamma$ case the value would not depart much from unity.} 
\label{fig:g2hgagacontourplot}
\end{center}
\end{figure}

As for the $\gamma \gamma$ and $Z \gamma$ decay modes we will only consider in this discussion the case of $h_0$, since the corresponding branching fractions are only relevant in the $120-160~\rm{GeV}$ mass range, where according to Ref.~\cite{Carone:2009nu} neither $\tilde{P}$ nor $\tilde{h}_0$ lie. In Fig.~\ref{fig:g2hgagacontourplot} we show contour plots of the absolute squared value of $g_{h_0 \gamma \gamma}$ and $g_{h_0 Z\gamma}$ in the $m_{h_0},m_{\tilde{h}_0}$ plane. For the case of $\gamma \gamma$, we see that the coupling squared varies by, at most, 10 \%. We have explicitly checked that in the region of parameter space not excluded from the analysis of Ref.~\cite{Carone:2009nu}, the actual value is always smaller than one. The $Z \gamma$ presents a variation of, at most, $1 \%$ in this same region.

From the previous paragraphs, concerning the neutral Higgs effective couplings, one can withdraw important information about the branching ratios.  It can be seen that from the different partial widths which scale with $g_{h_0gg}^2$, $ g^2_{h_0 VV}$, $g^2_{h_0 \gamma \gamma}$ and $g^2_{Z \gamma}$,  the $f \bar{f}$ and $gg$ modes scale with a larger factor than the other ones. Therefore, the branching fraction in the $f \bar{f}$ and $gg$ channels will be larger than in the SM, while the remaining will have a branching ratio lower than the SM value. Hence, the observable quantity production cross section times branching ratio (henceforth called \emph{rates}) can be, in principle, either suppressed or enhanced with respect to the SM and each rate should be studied separately.

The charged LWSM Higgs phenomenological analysis for the colliders search is not as appealing as the neutral one due to the conjunction of the Higgs sum rule in Eq.~\ref{eq:mhsumrule} and the LEP and Tevatron processes included in the {\tt {HiggsBounds}} routine. At LEP, one considers the process $e^+ e^- \to \tilde{h}^+ \tilde{h}^-$, where $m_{\tilde{h}^\pm} \le 94 ~\rm{GeV}$. At the Tevatron, the charged Higgs is produced via $t \to \tilde{h}^+ b$, with $\tilde{h}^\pm$ decaying into either $\tau \nu_{\tau}$ or $c \bar{s}$. For these analyses,  $m_{\tilde{h}^\pm} \le 155 ~\rm{GeV}$, and one needs the branching ratios of the top for the $\tilde{h}^{+} b$ and ${W}^{+} b$ channels. The region of parameter space being tested is $m_{h_0} \le 110~\rm{GeV}$, due to the Higgs masses sum rule. We note that, despite the aforementioned region is excluded by both the analysis of Ref.~\cite{Carone:2009nu} and the LEP bound on $m_{h_0}$, we still implement our charged Higgs sector into {\tt {HiggsBounds}}, in order to provide a cross-check on our results.

One of the main features of a light charged Higgs sector is the opening of a new decay channel for the top quark, $t\to h^+ b$.  In order to analyze the available experimental data on charged Higgs searches one needs the branching ratios of the top to the $h^+ b$ and $W^+ b$ channels.  In the LWSM it is easy to see that the former channel is only important for Higgs masses below $\sim 90$ GeV which is already ruled out by direct LEP searches \cite{:2001xy}.  

The charged Higgs decays can be divided into three groups, according to whether the decay is leptonic ($\tau^+ \nu_{\tau}, \mu^+ \nu_{\mu}$), hadronic ($u_i \bar{d}_j$) or into another Higgs boson plus a gauge boson.

The leptonic case is the easiest, since we are ignoring neutrino masses, and as such flavor mixing in the lepton sector. The partial width into this channel is given by
\be
\Gamma (H^{+} \to \ell^+ \nu_\ell )= \frac{G_F}{4 \sqrt{2} \pi} m_{\tilde{h}^+}^3 x (1-x)^2 \, .
\ee
where $\ell$ is the leptonic family index and $x=m_\ell^2/m^2_{\tilde{h}^+}$. 

The hadronic case is slightly different due to the presence of the CKM matrix and to the fact that both the up and down-type fermions have non-zero masses. The LO partial width is given by
\be
\Gamma (H^{+} \to u_i \bar{d}_j )= \frac{3 G_F}{4 \sqrt{2} \pi} \lvert V_{ij} \rvert^2 m_{\tilde{h}^+}^3 \sqrt{(1-x-y)^2-4xy} \left[ (1-x-y) (x+y) -4 xy \right] \, .
\ee
where $x=m^2_{u_i}/m^2_{\tilde{h}^+}, y= m^2_{d_j}/m^2_{\tilde{h}^+}$. The square root term comes from the $1 \to 2$ phase space integration, while the term in square brackets arises from the matrix elements squared.
It has been known since a long time \cite{Spira:1997dg} that in the above expression the bulk of the NLO corrections can be absorbed by using the QCD running masses instead of the pole masses.

The decay width of the third channel, which involves a charged Higgs and a $W$, is given by
\be
 \Gamma (\tilde{h}^{\pm} \to x W^{\pm})  = \frac{g_2^2}{64 \pi m_W^2 m^3_{\tilde{h}^\pm} }  \, g^2_{x \tilde{h}^{\pm} W^{\mp}}  \, \lambda^{3/2} (m_W,m_x,m_{\tilde{h}^{\pm}}) \, .
\ee

Before closing this section, we would like to stress that since the one-loop corrections to the production and decay of the SM Higgs boson, due to the Higgs boson itself, are known to be very small, and these corrections are not much different to those in the LWSM within the approximations used in this work, we expect our tree expressions to hold with sufficient accuracy at higher orders. 

\section{Numerical Results}
\label{sec:num}
In this section we use the above results in order to run the computer code {\tt {HiggsBounds}} (HB) to confront the LWSM Higgs sector against the available LEP and Tevatron data on Higgs search. 

The code HB, roughly speaking, uses the information from the previous section to compute, for each point in parameter space, the LWSM cross-section of the different Higgs search signal topologies.  Since HB is an exclusion code, the program chooses, for each point in parameter space, the LEP or Tevatron channel in which the ratio between the predicted LWSM cross-section and the background cross-section for the signal is maximized.  Therefore, this channel is the most sensitive in order to exclude the model in that point.  If the predicted LWSM cross-section of the selected process is greater than the observed cross-section then HB decides that the given point in parameter space is excluded at a $95\%$ C.L..  A detailed explanation on the HB code may be found elsewhere \cite{Bechtle:2008jh,Bechtle:2009ic,Bechtle:2009ni,Bechtle:2010jt,Bechtle:2011sb}.

Since HB works under the narrow width approximation, there are two requirements to be fulfilled by the points in parameter space to be tested.  On one hand, all Higgs bosons mass and width shall fulfill the narrow width approximation, which technically limits its applicability to Higgs bosons not heavier than $\sim 600~\rm{GeV}$.  This upper value, which is the highest value quoted in the experimental MC studies of both ATLAS and CMS, gives a width to mass ratio $\sim 0.2$.   On the other hand, in order to have the signals disentangled, the Higgs bosons have to be separated in mass much more than the maximum of their width.  In the LWSM we find that this condition is fulfilled if $m_{h_0} \le 0.9 m_{\tilde{h}_0}$. This requirement automatically satisfies the non-perturbativity constraint of $y_t^2 \le 4 \pi$.

A third requirement to run HB is that the model does not change the expected background substantially, which is accomplished in the LWSM.

\subsection{Impact of collider bounds: LEP and Tevatron}

We will start by examining the impact of the direct searches of LW Higgs boson. Along this subsection we will not include the constraints on the charged Higgs sector, just to assess the coverage of the existing exclusions from both LEP and Tevatron.

\begin{figure}[htbp]
\begin{center}
 \begin{minipage}[b]{0.45\linewidth}
\begin{center}
\includegraphics[width=1\textwidth]{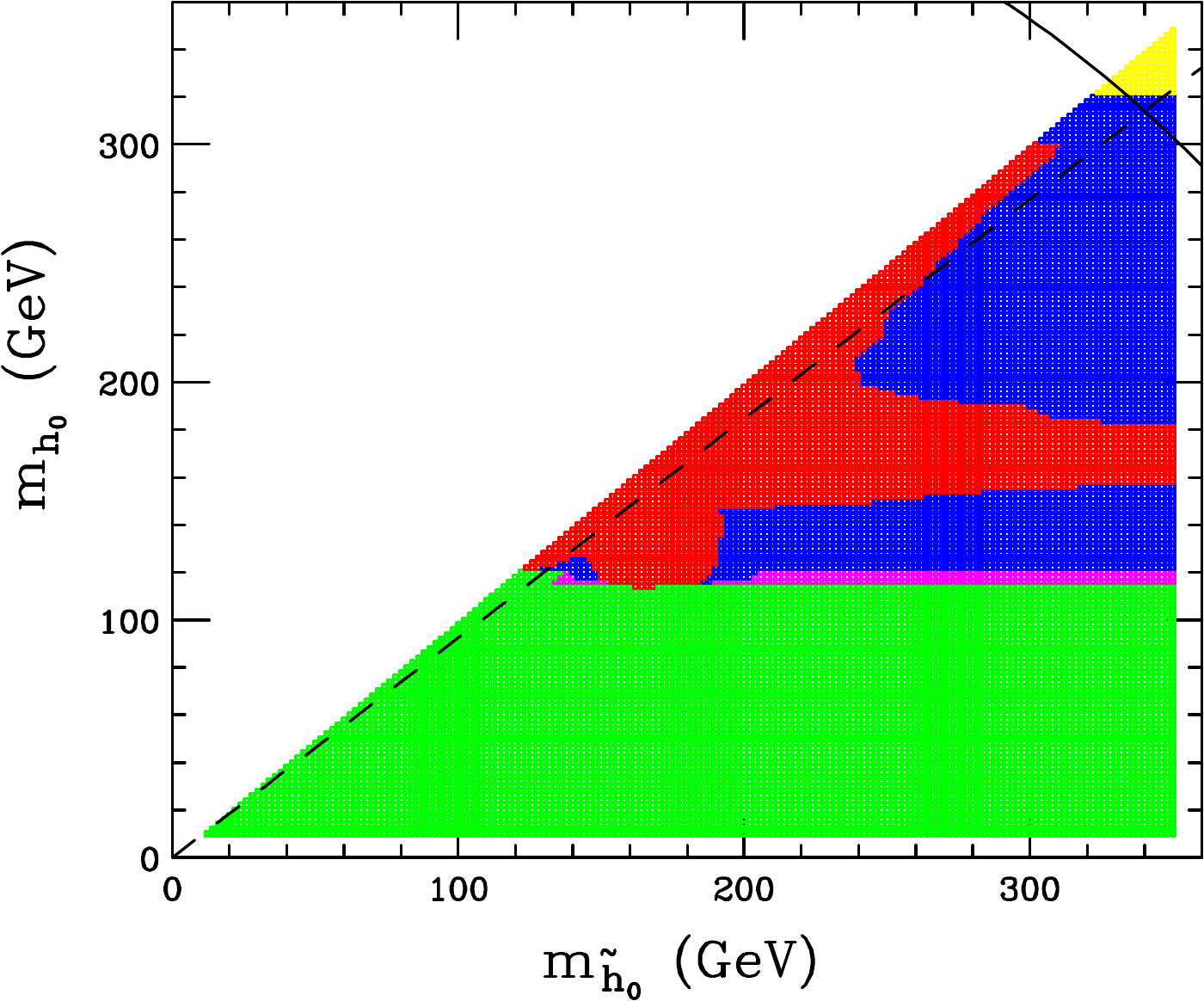}
\newline
(a)
\end{center}
\end{minipage}
\hspace{0.5cm}
\begin{minipage}[b]{0.45\linewidth}
\begin{center}
\includegraphics[width=1.02\textwidth]{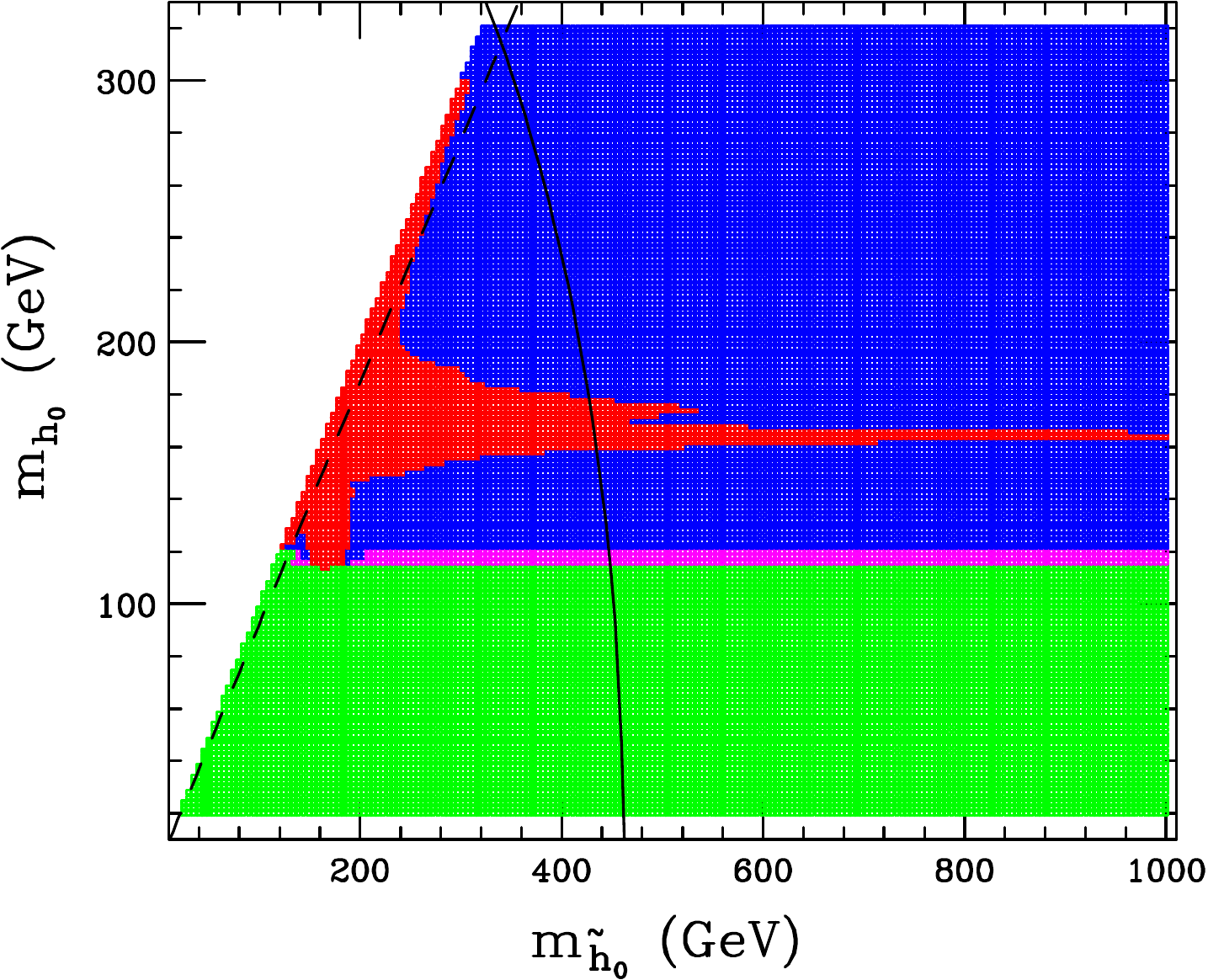}
\newline
(b)
\end{center}
\end{minipage}
\caption{ $m_{h_0}$ as a function of $m_{\tilde{h_0}}$, obtained with {\tt {HiggsBounds}} 2.0.0. The green (red) points are excluded by LEP (Tevatron), and the magenta (blue) are those where the most sensitive channel comes from LEP (Tevatron) data. Yellow points are out of current collider reach. The region to the left of the black line are the points excluded by the analysis of Ref.~\cite{Carone:2009nu}. The top-Yukawa coupling squared is less (greater) than $4 \pi$ in the below (above) the dashed line. }
\label{fig:4}
\end{center}
\end{figure}

In Fig.~\ref{fig:4} we show the collider bounds on the model, in the $m_{h_0}, m_{\tilde{h}_0}$ plane. We employ the following color code: points in green (red) are excluded by LEP (Tevatron), and the magenta (blue) are those where the most sensitive channel comes from LEP (Tevatron) data. In the left panel, we select a mass range where both $h_0$ and $\tilde{h_0}$ can be excluded. In the right panel, we extend the range of $m_{\tilde{h}_0}$ up to $1~\rm{TeV}$. The SM reference values are taken from the internal subroutines of {\tt{HiggsBounds}}.
%Here, a word of caution is necessary: as noted in Ref.~\cite{Chivukula:2010nw}, a heavy Higgs will generate a negative correction to the T parameter, and LW fermions, due to its extra minus sign, can not compensate this effect. In the next subsection, we will address the impact of the EWPO in the LW Higgs sector, but here we will include only the collider bounds from direct searches.
The region to the left of the solid line is excluded by the analysis of Ref.~\cite{Carone:2009nu}, and in the one above the dashed line the top Yukawa coupling becomes non-perturbative. The inclusion of the newer version of {\tt {HiggsBounds}} excludes a significant portion of the parameter space, due to the fact that the $gg \to H \to WW$ exclusion is stronger, since it incorporates the most recent data \cite{Aaltonen:2010sv}. This study does not only has a stronger bound due to the increased luminosity, but also extends the published kinematical range of the exclusion from $200~ \rm{GeV}$ to  $300~ \rm{GeV}$, and includes other analyses,  that can probe Higgs masses up to $320 ~\rm{GeV}$.

As we have predicted, the LEP bound of $114.4~\rm{GeV}$ (green points) holds over all the parameter space. Actually, we have checked that the exclusion of the green points is only due to the $e^+ e^- \to h Z, \, h \to b \bar{b}$. For high values of $r$, where both CP-even Higgs bosons are close in mass, one finds that the exclusion can reach up to $120~\rm{GeV}$, which is the maximum mass value published by the LEP collaboration in this channel. As for the Tevatron exclusions, they come entirely from the latest dedicated study of the $g g \to h_0 / \tilde{h}_0 \to WW$ channel\cite{Aaltonen:2010sv}, where both CDF and D0 results have been combined, with each experiment contributing with a total luminosity of 4.8 and 5.4 $fb^{-1}$ respectively. It is clear that the horizontal stripe around $m_{h_0} \sim 165~\rm{GeV}$ correspond to the exclusion due to $h_0$. What might seem slighlty less clear, is that the vertical stripe around $m_{\tilde h _0} \sim 165~\rm{GeV}$ correspond to the same experimental search, but this time with $\tilde{h}_0$. The blue regions around those stripes do also make sense: they correspond to those points where the reach in the $h \to WW$ channel is not enough to exclude points, but, however, it is still more sensitive than LEP and other Tevatron searches. The horizontal exclusion band becomes narrower as soon as $r$ decreases (or, equivalently, the mass of $\tilde{h}_0$ increases), since the enhancements on the couplings tend to be smaller.  On the other hand, the vertical exclusion band becomes much narrower as $r$ decreases (or the mass of $h_0$ decreases), since in this case $g_{\tilde h_0 VV}\to 0$.  We stress the fact that the collider bounds are able to exclude a portion of the parameter space not probed before by the constraints coming from Ref.~\cite{Carone:2009nu}. Let us note that we can make the $m_{\tilde h_0}$-independent exclusion of the $163-166 ~\rm{GeV}$ range for $m_{h_0}$, which comes from analyzing the large $m_{\tilde h_0}$ limit, while for the lower allowed values of  $m_{\tilde h_0}$, the exclusion covers the $160-175 ~\rm{GeV}$ range.

It is worth noticing that this limit, which should correspond to the SM case, does not retrieve the well known CDF and D0 combined analysis \cite{Aaltonen:2011gs} which excludes the $158-173$ GeV region, because {\tt {HiggsBounds}} does not combine analysis. This is due to the fact that {\tt {HiggsBounds}} works on a channel-by-channel basis, and therefore it does not perform any combination of channels for the same Higgs (it does, however, combine channels if they correspond to different Higgs bosons).
%As a matter of fact {\tt {HiggsBounds}} only performs isolated analysis. 
  It is expectable, given the increasing behaviour of the effective couplings $g_{h_0 VV}$ and $g_{h_0 f \bar f}$, that a full combined analysis of the LWSM Higgs Sector would exclude a much larger region of $h_0$ masses as far as $m_{\tilde h_0}$ is not too large.     
\subsection{Impact of collider bounds: projection and perspectives for LHC}\label{sec:LHC}

We analyze in this section the potential discovery of LHC for the LW Higgs sector.  In particular we focus on a SM-like Higgs boson with mass between $110$ GeV and $200$ GeV and we restrict to points with $m_{\tilde h_0}$ below $1$ TeV and not excluded by the $b\to s \gamma$ ($420$ GeV$<m_{\tilde h_0}<1000$ GeV). These considerations restrict the range of $r$ to the interval $0.11 - 0.48$. 

\begin{figure}[!htbp]
\begin{center}
\begin{minipage}[b]{0.45\linewidth}
\begin{center}
\includegraphics[width=1\textwidth]{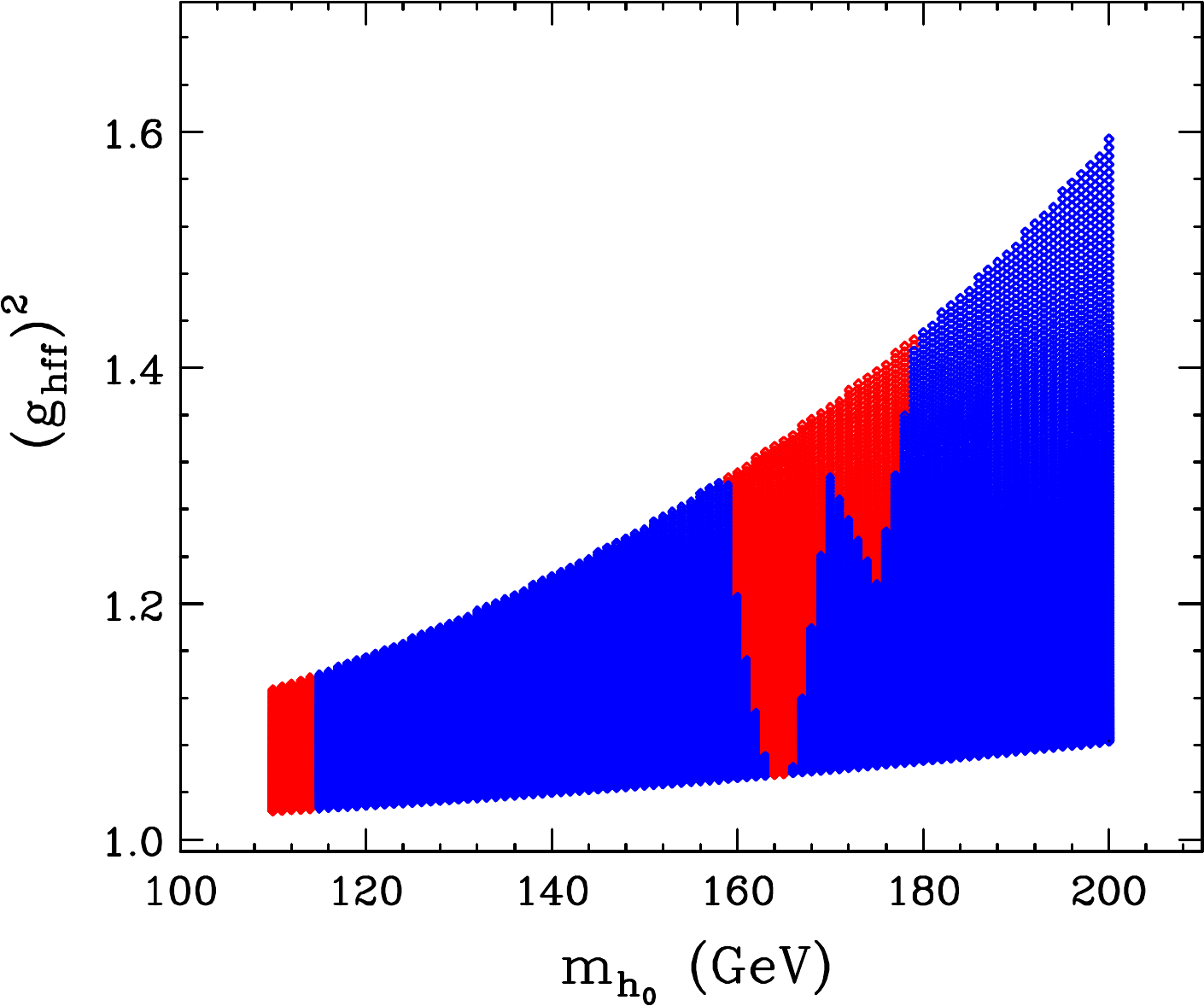}
\newline
(a)
\end{center}
\end{minipage}
\hspace{0.5cm}
\begin{minipage}[b]{0.45\linewidth}
\begin{center}
\includegraphics[width=1\textwidth]{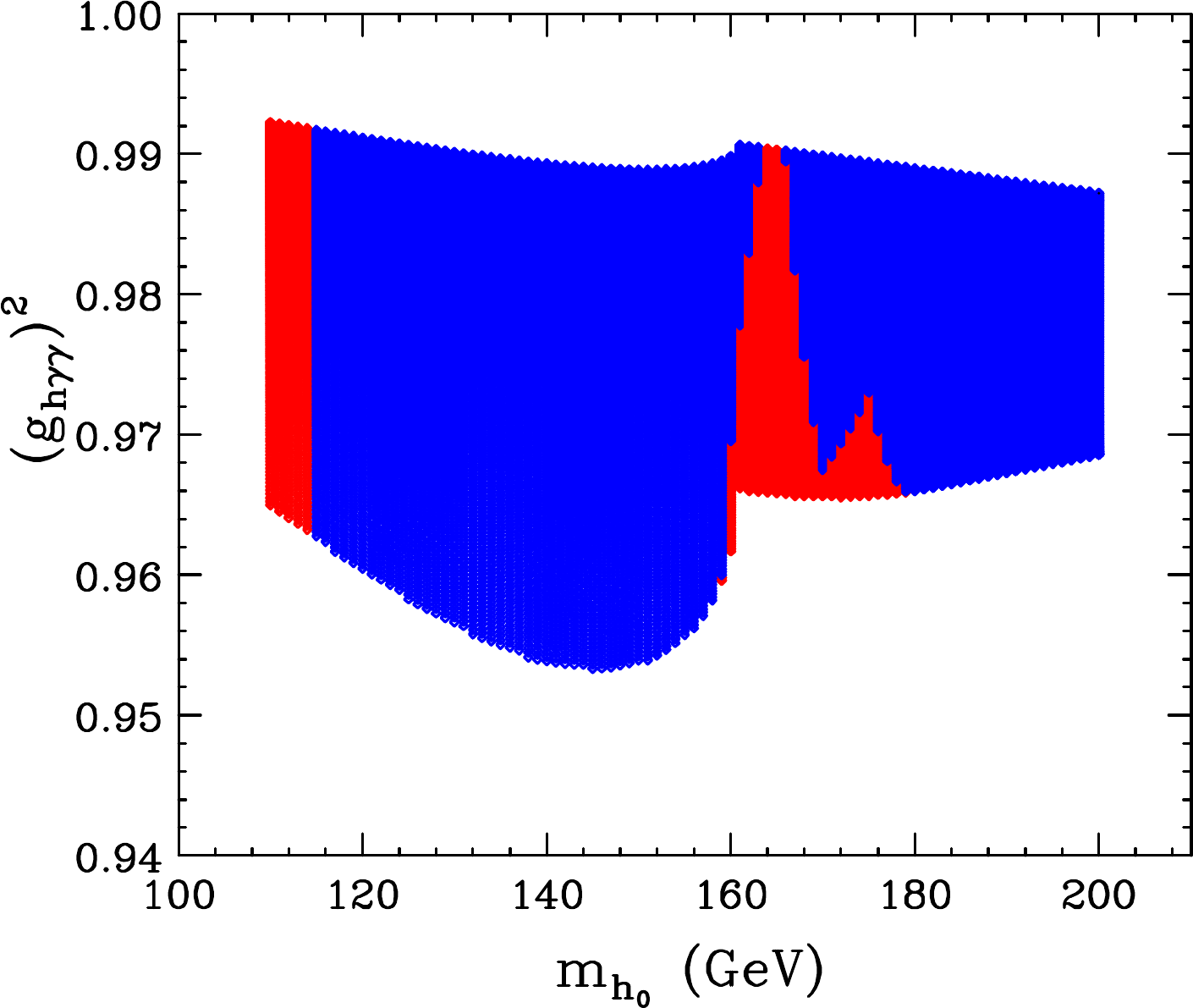}
\newline
(b)
\end{center}
\end{minipage}
\caption{$ g^2_{h_{0} f \bar{f}} $ (a) and $ g^2_{h_{0} \gamma \gamma} $ (b) as a function of $m_{h_0}$, obtained with {\tt {HiggsBounds}} 2.1.1. We show only the points not constrained by the analysis of Ref.~\cite{Carone:2009nu}. The red (blue) points are excluded (allowed) by present collider data. }
\label{fig:5}
\end{center}
\end{figure}

At this point it is worth looking at the effective couplings again, now restricted to this region of parameter space, in order to asses the impact on LHC phenomenology. In Fig.~\ref{fig:5} we show the coupling of the lightest Higgs boson to fermions and photons.  The increase in $g_{h_0 f \bar{f}}$ is at most of $30 \%$ for Higgs masses below $160~\rm{GeV}$, and can go up to $60 \%$ for Higgs masses between $180$ and $200~\rm{GeV}$. As discussed in subsection \ref{subsec:phenoanal} this constitutes an upper bound for the enhancement in the total cross section, and also for the rates in the channels involving electroweak gauge bosons, due to the reduction of the branching fractions. From the right panel we see that the coupling to photons is reduced with respect to the SM. However, this reduction is not dramatical: at most, 5 \%.

 We stress the fact that, since one has to recover the SM result in the $r \to 0$ limit, the points where higher deviations from the SM occurs correspond to higher values of $r$, or, correspondingly, to a not too heavy $\tilde{h}^0$ boson.  In this region we may expect to be able to differentiate the LWSM from the SM.
 
We note that the relative coupling of the CP-even Higgs bosons to EW gauge bosons (not shown here) presents a similar shape to the fermionic one, but its value (as mentioned before) is always below than the one for fermions for the same Higgs mass. More concretely, looking into the currently not excluded points, for $g_{h_0 VV}^2$ we find a maximum value of $1.02~(1.06)$  for $m_{h_0}$ in the $110-160~\rm{GeV}~(180-200~\rm{GeV})$ , and for $g_{\tilde{h}_0 VV}^2$ one has $0.02~(0.06)$ in the same range of masses. Evidently, in this region the heavy CP-even Higgs in gaugephobic, and thus the scenario where both CP-even Higgs couple stronger than the SM Higgs to gauge bosons is not achievable: $g_{\tilde h_0 VV}^2 > 1$ only if $r > 0.84$.

Having defined the range to scan in the $m_{h_0}, m_{\tilde{h}_0}$ plane, and briefly analyzed our expectations for the rates, now we move into the study of the LW Higgs sector at the LHC. Here we focus on the Run I (until the end of 2012), using the expected exclusion limits provided by the ATLAS collaborations \cite{Atlas:hgaganote,Atlas:HiggsNote}.\footnote{ The choice of the ATLAS studies over the corresponding ones from CMS \cite{CMS:HiggsNote} is only due to their more stringent exclusion limits.}
The aforementioned studies include the $\gamma \gamma$ and $WW$ decay modes for masses below $200~\rm{GeV}$, and also the production of the Higgs by vector boson fusion and the decay into either bottoms or taus, in the same mass range.  For heavy Higgs bosons (with masses below $600~\rm{GeV})$ they also include the  exclusion limits in the $ZZ$ channel. The SM reference values for cross sections and branching ratios at the LHC are obtained from the LHC Higgs Working Group report \cite{LHCHiggsCrossSectionWorkingGroup:2011ti}.

\begin{figure}[!htbp]
\begin{center}
\begin{minipage}[b]{0.45\linewidth}
\begin{center}
\includegraphics[width=1\textwidth]{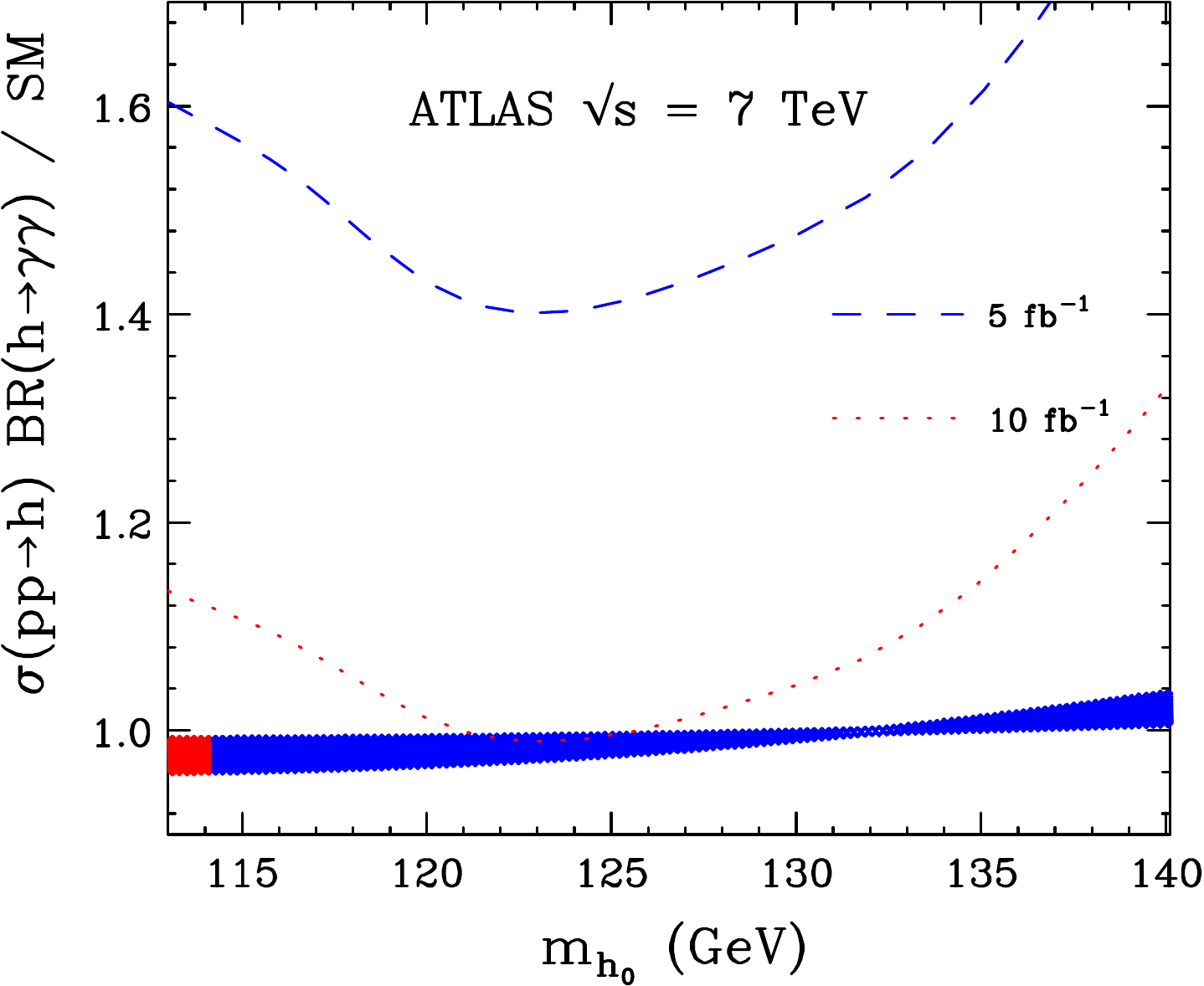}
\newline
(a)
\end{center}
\end{minipage}
\hspace{0.5cm}
\begin{minipage}[b]{0.45\linewidth}
\begin{center}
\includegraphics[width=1\textwidth]{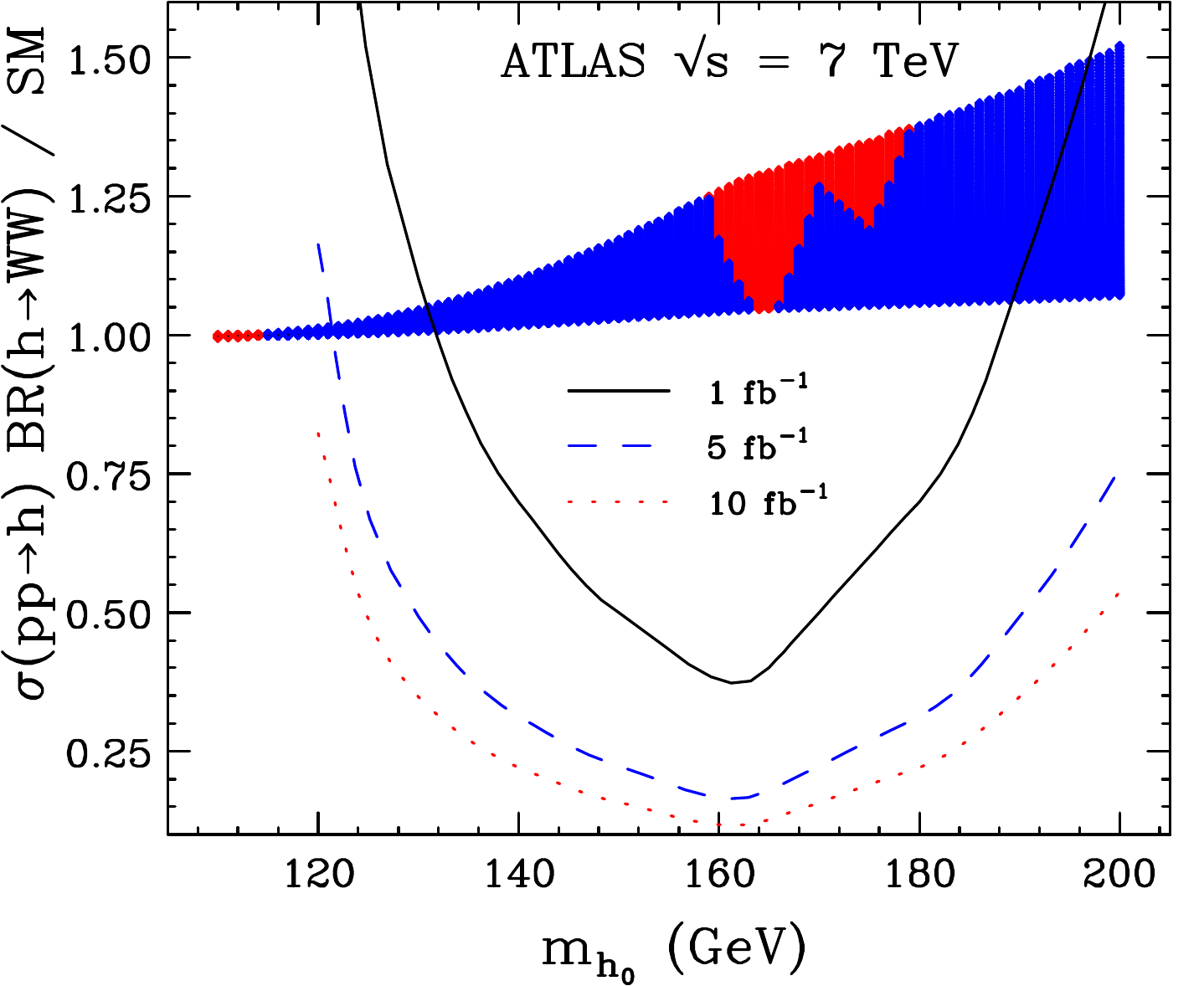}
\newline
(b)
\end{center}
\end{minipage}
\caption{Total cross section times branching ratio of the lightest CP-even Higgs boson $h_0$ in the (a) $\gamma \gamma$ and (b) $WW$ channels. The red points are excluded by current collider data, while the blue points are not. We only show here points satisfying the $b \to s \gamma$ constraint. The LHC reach is shown for three different integrated luminosities: 1 $\rm{fb}^-1$ (solid line), 5 $\rm{fb}^-1$ (dashes) and 10 $\rm{fb}^-1$ (dots)}
\label{fig:htoVV}
\end{center}
\end{figure}

In Fig.~\ref{fig:htoVV} we show the cross section times branching ratio for the decay modes of the lightest CP-even Higgs into $WW$ and $\gamma \gamma$ as a function of the Higgs mass,  at the LHC and for a center-of-mass energy of 7 TeV. We also show the reach of each channel and assume three different integrated luminosities: 1 (end of 2011), 5 (end of 2012, realistic) and 10 $fb^{-1}$ (end of 2012, optimistic). The exclusion limits for the reference luminosity of $1~\rm{fb}^{-1}$ are taken from Refs. \cite{Atlas:hgaganote,Atlas:HiggsNote}, and for the other two scenarios, we scale the expected result by the square root of the luminosity: this procedure yields the right result if one assumes a background dominated regime and neglects all systematics errors.
% PLOT EXTRA # 1: signals in WW and ZZ for light Higgs.

From the figure we see that the $\gamma \gamma$ channel has actually a rate lower than the SM for $m_{h_0} \le 128~\rm{GeV}$. For masses below $140~\rm{GeV}$ the difference between the SM and LWSM rates is at most 5 \%. This means that, at the LHC, it would be impossible to distinguish this scenario from the SM in the diphoton channel, since in the $120-140~\rm{GeV}$ range a change of at least $20 \%$ is necessary \cite{Duhrssen:2004cv}. 

As for the gauge boson channel, we see that the increase is of at most $30\%$ (60 \%) for a $160\ (200)~\rm{GeV}$ Higgs. In this case one can distinguish the LW Higgs from the SM if the enhancement is larger than 10 \%, with $300~\rm{fb}^{-1}$ of data \cite{duehrssennote}.

From Figure \ref{fig:htoVV} it can be seen that, with $1~\rm{fb}^{-1}$ of data, the $WW$ channel would already be able to exclude most of the points, except for bosons with masses above $\sim 180~\rm{GeV}$, or below $130~\rm{GeV}$. For the heavy mass range, the $ZZ$ decay mode can exclude an important fraction of points already with this luminosity, and for higher luminosities both channels completely cover this area. 
For an integrated luminosity of 5 fb$^{-1}$, one probes masses larger than 122 GeV, while in the optimistic scenario of 10 fb $^{-1}$ the exclusion extends to $120~\rm{GeV}$ masses.

The $120~\rm{GeV}$ limitation is simply due to the fact that both CMS \cite{CMS:HiggsNote} and ATLAS \cite{Atlas:HiggsNote} cut off the analysis of these channels at 120 GeV.  However, the recent update of ATLAS for the SM Higgs boson sensitivities \cite{Atlas:SMHiggsNote}, where different channels are combined, reports that a total luminosity of  $4.2 \rm{fb}^{-1}$ is required in order to exclude a 115 GeV SM Higgs. Therefore it is rather likely, that a combination of all possible channels (and maybe also a combination of both collaborations) would be able to fully test the lightest LW Higgs boson at the LHC Run I, even without assuming the optimistic 10 fb$^{-1}$ scenario.

Now we briefly comment on the prospects for the other LW Higgs bosons. In increasing order of its masses, the next boson is $\tilde{h}_0$. Since, as we stated before, this boson is gaugephobic, and  its mass is above the top pair production threshold (due to the masses sum rule), we find the $t\bar t$ channel to be its main decay mode with a branching fraction always higher than $60 \%$. The second decay mode is actually the $\tilde{h}_0 \to h_0 h_0$, whose branching fraction varies, roughly speaking, in the $1-30$ \% range. This provides a very interesting decay mode of either four fermions, or two fermions plus two photons; however this signature will only be at the reach of LHC with very high luminosities (see Ref.~\cite{Buescher:2005re}).

As for the remaining Higgs bosons, we find that the charged Higgs decays into top-bottom with a ratio larger than 0.95. Together with the $\tilde{h}^\pm \to h_0 W^{\pm}$ channel, they comprise the two observable decay channels of the charged Higgs, since all the others have branching fractions below $10^{-5}$. The CP-odd Higss $\tilde{P}$ decays predominantly into tops; the $\tilde{P} \to h Z$ channel account for, at most, $10 \%$, and the gluon-gluon and $b \bar{b}$ channels account for a 1 \% and 0.1 \% of the decays, respectively. It is therefore evident that to directly probe these heavy Higgs bosons at the LHC high luminosities are required: these channels also appear in the MSSM, and typical studies are done with, at least, 30 fb$^{-1}$ \cite{Buescher:2005re}.

To summarize this subsection, we have found that the lightest LW Higgs boson can be excluded at the LHC Run I, while the existence of other Higgs bosons can not be directly probed. If a LW Higgs signal is seen at Run I, it will look like a SM Higgs: a very high luminosity would be required to rule out the SM case, and it might indeed not be enough\cite{duehrssennote,Buescher:2005re}. If the measure of the $WW$ channel is higher than the SM cases, then one expects the remaining Higgs bosons to be not too heavy. If the $WW$ channel measurement is compatible with the SM Higgs, then that could mean that the remaining LW Higgs bosons are heavier. In either case, one would still need to directly probe the other Higgs bosons. We have also found that the decay modes of the heavy Higgs bosons is rather restricted, mainly due to the hierarchy in the Higgs spectrum that does not allow, for instance, one of the heavier Higgs bosons to decay into another one plus a gauge boson ( interesting decays like $\tilde{h}^\pm \to W^{\pm} \tilde{h}_0$, or $\tilde{P} \to Z \tilde{h}_0$ are kinematically forbidden).

\section{Conclusions}\label{sec:conclu}

In this work we have studied the bounds on the LW Higgs sector coming from direct searches at both the LEP and Tevatron collider, using the code {\tt {HiggsBounds}}. We have also analyzed the prospects for direct detection at the LHC Run I. This work complements the study of the indirect constraints performed by Carone and Primulando~\cite{Carone:2009nu} and settles down the state of the art and perspectives for direct search of LW Higgs Bosons in colliders.

While direct collider searches place strong constraints mostly on $m_{h_0}$, the observables considered in Ref.~\cite{Carone:2009nu} are mostly sensitive to the loop contribution of the charged Higgs boson. These indirect constraints rule out a great portion of the parameter space ($m_{\widetilde{h}^{\pm}} > 463~\rm{GeV}$), while we have shown that the current Tevatron result is only able to constraint a small region of parameter space.  A light SM-like LW Higgs boson with a mass in the $163-166~\rm{GeV}$ range is forbidden (independently of the non SM-like LW mass scale), as can be seen from Fig.~\ref{fig:4}. If $\tilde{h_0}$ is not too heavy ($\sim 450~\rm{GeV}$) then the exclusion for $m_{h_0}$ extends to the $150-175~\rm{GeV}$.
However, by the same token, as soon as the LHC starts to probe a wider range of masses for $h_0$, the exclusion will become stronger, as shown in Fig.~\ref{fig:htoVV}: already with $1~\rm{fb}^{-1}$ of data $h_0$ masses between $130$ and $180~\rm{GeV}$ can be ruled out.

In this model, all the production modes of $h_0$ are always enhanced with respect to the SM, and also its branching fraction into fermions. Conversely, the decay modes involving gauge bosons ($ZZ, WW, Z \gamma, \gamma \gamma$) are always reduced with respect to the SM case. When analyzing the rates for the $WW$ and $\gamma \gamma$ channel, we have found that in the $WW$ channel one always obtains an enhancement of the total rate,  (it can go up to 50 \% for $m_{h_0} \sim 200~\rm{GeV}$), while in the photon channel suffers a slight reduction (at most, 5 \%) in the interesting light mass range ($m_{h_0} \le 128~\rm{GeV})$, and it can be enhanced by at most 4 \%, for heavier masses.

The magnitude of these two enhancements provides important messages. On one hand, the $WW$ channel enhancement means that the LHC will be able to test a wider range of masses for $h_0$ than for the SM Higgs. With only $1~\rm{fb}^{-1}$ of data, it would be already able to probe a significant portion of parameter space, which can be further extended into the lighter mass region with increased luminosity. Taking into account the current projections for the SM Higgs, one can state that the $h_0$ can be fully probed by the LHC Run I. Moreover, if $h_0$ is not too light, the LHC Run II (around $300~\rm{fb}^{-1}$ of data) will be able to tell apart the LW scenario from the SM. The mild variation (with respect to the SM) of the $\gamma \gamma$ channel does also have phenomenological implications for LHC Run I. On one side, this channel alone is not a discovery channel for LHC Run I, not even if a total integrated luminosity of $10~\rm{fb}^{-1}$ (optimistic scenario) were to be achieved. Moreover, the LHC Run II will not be able to use the diphoton channel to tell apart $h_0$ from the SM Higgs.

One important feature of the LW Higgs sector, is the fact that both CP-even Higgs bosons can couple stronger to the $W$ and $Z$ than the SM Higgs. While this is a very appealing possibility, we have also shown that, in its minimal realization, this scenario does not occur. However, a non-minimal LW Higgs sector might be able to accommodate this interesting phenomenological possibility.

As for the remaining Higgs bosons, the main decay channel is $t \bar{t}$ (top bottom) for the neutral (charged) bosons. However, for the heavy CP-even Higgs, the detection through $\tilde{h}_0 \to h_0 h_0 \to \gamma \gamma b \bar{b}$ constitutes a promising channel, and it might be worth fully exploring the consequences of this decay mode. Similar Higgs chain decays modes do not occur with a substantial rate, mainly due to kinematically closure of these channels, due to the strict hierarchy in the Higgs spectrum. It would also be interesting to analyze if radiative corrections can alter the spectrum hierarchy.

\begin{acknowledgements}
We thank Marc Gillioz and Alejandro Szynkman for reading the manuscript.  We also thank Roman Zwicky for useful discussions. The work of J.Z.~is
supported by the Swiss National Science Foundation (SNF) under Contract No. 200020-126691.
\end{acknowledgements}

\begin{appendix}

\section{Loop functions}\label{app:loopfunc}

In this appendix we collect all of the relevant formulae for the loop functions that appear in Subsection \ref{subsec:loopdec}. The expressions are adapted from \cite{Gunion:1989we}.

For $h_0,\,\tilde h_0,\,\tilde P \to \gamma \gamma$ one defines the function $f(x)$ as
\bea\label{eq:func}
f(x) = \left\{ 
\begin{array}{ll}
\arcsin^2(1/\sqrt{x})  & \qquad \text{if}  \, x \ge 1   \\
 -\frac{1}{4} [\log ( \frac{1+\sqrt{1-x}}{1-\sqrt{1-x}}) - i \pi]^2 & \qquad \text{if} \, x < 1    
\end{array} \,\right. .
\eea
We recall that $f(x) \approx \frac{1}{x} + \frac{1}{3 x^2} $ in the $x \to \infty$ limit.
The loop functions for bosons ($1$) and scalars ($0$) for the $\gamma \gamma$ case are simply given by\footnote{Notice that $F_0$ differs from $F_0^{\eta}$ of Ref.~\cite{Krauss:2007bz} by an overall minus sign.}
\bea
F_1 (x) &=& 2 + 3 x + 3 x (2-x) f(x) \, ,\\
F_{0} (x) &=& x (1-x f(x) )  \, .
\eea
For the fermions ($1/2$) one has to distinguish between CP-even and CP-odd Higgs bosons. As such, we define
\bea
F_{1/2} (x) &=&-2 x (1+ (1-x) f(x) ) \, , \\
F^{\tilde{P}}_{1/2} (x) &=& -2 x f(x) \, ,
\eea
where $F^{\tilde{P}}_{1/2}$ is the function for the $\tilde{P}$, and, for simplicity, we omit the superscript in $F_{1/2}$ for the case of the CP-even Higgs bosons. Note that Eq.~(\ref{eq:Pgg}) is derived simply by taking the absolute squared value of the ratio of $F^{\tilde{P}}_{1/2}$ and $F_{1/2}$.

For the $h_0,\,\tilde h_0,\,\tilde P \to \gamma Z$ decays, one has to define the following functions,
\bea
A_t^x  ( \beta_x^t,\beta_Z^t)&=& N_c \frac{(-2Q_t)}{s_W c_W } g_V^{t} \bigl(\zeta_x I_1( \beta_x^t,\beta_Z^t) - I_2  ( \beta_x^t,\beta_Z^t) \bigr) \, ,  \\
A_W (\beta_x^W, \beta_Z^W) &=& - t_W^{-1} \bigl\{ 4 (3-t^2_W) I_2  ( \beta_x^W,\beta_Z^W) +\nonumber \\
&& \bigl[ (1+\frac{2}{\beta_x^W}) t_W^2- (5+ \frac{2}{\beta_x^W})\bigr]  I_1  ( \beta_x^W,\beta_Z^W) \bigr\} \, ,\\
A_{\tilde h^{\pm}} (\beta_x^{h^{\pm}},\beta_Z^{h^{\pm}}) &=& \frac{1-2 s^2_W}{c_W s_W} I_1 (\beta_x^{h^{\pm}},\beta_Z^{h^{\pm}}) \frac{m_W^2}{m_{\tilde h^\pm}^2}\, , 
\eea
where $s_W=\sin \theta_W, c_W= \cos \theta_W, t_W=\tan \theta_W$, $Q_t=2/3$, $N_c=3$, $g_V^{t}=I^{3 L}_t - 2 Q_t s^2_W$ is the vectorial coupling of the top quark and $\zeta_x$ is 1 (0) for CP-even (odd) Higgs bosons.  The index $x$ refers to which Higgs ($h_0,\ \tilde h_0$ or $\tilde P$) is the effective coupling sought. 
The functions $I_{1,2}$ read% (Eqs. (2.24) and (2.25) of Ref. \cite{Gunion:1989we})
\bea
I_1 (a,b) &=& \frac{ab}{2 (a-b)} + \frac{a^2b^2}{2 (a-b)^2} [f(a)-f(b)] + \frac{a^2b}{(a-b)^2} [g(a)-g(b)] \, ,\\
I_2 (a,b) &=& - \frac{ab}{2 (a-b)} [f(a)-f(b)] \, ,
\eea
with $f$ given by Eq.~(\ref{eq:func})  and $g$  by 
\bea\label{eq:gfunc}
g(x) =  \left\{
\begin{array}{cc}
\sqrt{x -1 }\arcsin (1/\sqrt{x})  & \qquad \text{if}  \, x \ge 1   \\
 \frac{1}{2} \sqrt{1-x} \bigl[ \log ( \frac{1+\sqrt{1-x}}{1-\sqrt{1-x}}) - i \pi \bigr] & \qquad \text{if} \, x < 1    
\end{array} \right. \, .
\eea

\end{appendix}

%  \newpage

  \end{document}